\documentclass[aps,pre,amssymb,amsmath,twocolumn,floatfix]{revtex4-2}
\usepackage{graphicx}
\usepackage{amsmath}
\usepackage{xcolor}

\usepackage{multirow}
\usepackage[left=2cm,right=1cm,top=2cm,bottom=2cm]{geometry}

\begin{document}
\title{Phase probabilities in first-order transitions using machine learning}
\author{Diana Sukhoverkhova$^{1,2}$}
\author{Vyacheslav Mozolenko$^{1,2}$}
\author{Lev Shchur$^{1, 2}$}
\affiliation{$^1$ Landau Institute for Theoretical Physics, 142432 Chernogolovka, Russia}
\affiliation{$^2$ HSE University, 101000 Moscow, Russia}

\begin{abstract}
We set out to explore the possibility of investigating the critical behavior of systems with first-order phase transition using deep machine learning. We propose a machine learning protocol with ternary classification of instantaneous spin configurations using known values of disordered phase energy and ordered phase energy. The trained neural network is used to predict whether a given sample belong to one or  another phase of matter. This allows us to estimate for the first time the probability that configurations with a certain energy belong to the ordered phase, coexistence phase, and disordered phase.  Based on these probabilities, we obtained estimates of the values of the critical energies and latent heat  for the Potts model with 10 and 20 components, which undergoes a strong discontinuous transition. We also found that the probabilities  may reflect geometric transitions in the coexistence phase.  
\end{abstract}

\maketitle

\section{Introduction} 

The application of deep neural networks for supervised machine learning~\cite{Sen-2020} to study the critical behavior of models with second-order phase transition allows us to estimate the critical temperature~\cite{Carrasquilla-2017,Carleo-2019} and the critical exponent of the correlation length~\cite{Chertenkov-2023}. Examples include the study of Ising model, the Baxter-Wu model, the Potts model and the XY model, and  percolation problem and many others~\cite{Carrasquilla-2017,Van Nieuwenburg-2017,Morningstar-2018,Westerhout-2020,Deng-2022,Bachtis-2020,Miyajima-2023,Chertenkov-2023}. The approach proved to be quite robust also in the study of the Ising model with non-trivial diagonal anisotropy~\cite{SS-jetpletters}  and in cross-training between universality classes~\cite{CS-prl}. This approach is based on binary classification of Monte Carlo-generated instantaneous configurations of the model into ferromagnetic and paramagnetic phases during training and the application of a trained neural network to predict whether the tested instantaneous configurations generated at a known temperature belong to one of these two phases. In this way, the probability distribution in sample space at temperature $T$ of belonging to the ferromagnetic or paramagnetic phase is estimated.  Finite-size analysis of this function and its variation allows us to estimate with satisfactory accuracy the critical temperature and the exponent of the critical correlation length~\cite{Carrasquilla-2017,Chertenkov-2023}. 

In the case of phase transitions of the first order, the phase transition temperature can also be estimated by a learning/testing approach similar to the one mentioned above, using learning relative to a known critical temperature. However, with this approach, it is not possible to estimate the values of the critical energies and hence the magnitude of the latent heat, i.e., the difference between the energies of the ordered $e_o$ and disordered $e_d$ phases at the phase transition temperature. A neural network trained on binary classification cannot capture the coexistence phase, which is a hallmark of systems with phase transition of the first order. A different approach is required.  In this letter, we propose a new method for solving such a problem.

The method is based on supervised learning, but instead of binary classification, a ternary classification of instantaneous spin configurations is used.  The classification is performed relative to known critical values of energies $e_o$ and $e_d$: OS - ordered phase for samples with energy $e<e_o$, CS - coexistence phase for samples with energy $e_o<e<e_d$, DS - disordered phase for samples with energy $e>e_d$~\cite{footnote}. During testing, the spin configuration snapshot obtained at a specific energy $e$ is fed to the input of the neural network, and the network produces three numbers corresponding to predictions that the tested configuration with energy $e$ may belong to one of the three phases. Based on testing a large number of configurations at the same value of energy $e$, we obtain an estimate of the probability that the tested snapshots with energy $e$ belong to one of the three phases. 
 
The application of such a method requires a large number of uncorrelated sample data sets with a certain energy value for training and testing the neural network. Modeling such datasets usually takes a large amount of time~\cite{Neuhaus-2003,Mayor-2007}. Fortunately, a microcanonical population annealing (MCPA)~\cite{Rose-2019,MS-2024} algorithm has recently been developed that generates a large number of replicas of the system under study using parallel acceleration on GPUs and filters a fraction of the replicas at a given energy~\cite{Rose-2019,MS-2024}. We simulated $2^{17}$ replicas for the Potts model with 10 and 20 components~\cite{Potts-1952}. Detailed analysis showed good qualities of the method both in comparison with another microcanonical Wang-Landau method and with known exact results~\cite{Fadeeva-2024}.  In the simulation, at each step of the algorithm, we randomly selected at each energy value $2^{13}=8192$ replicas from the current replica pool and used them for training the neural network and for testing and analysis. The density of states in the neighborhood of critical energies is a decreasing function of energy, and when applying the MCPA ceiling algorithm step with decreasing energy~\cite{Rose-2019} most of the population of $2^{17}$ configurations will be concentrated near the energy ceiling.  Furthermore, all replicas are randomly ``equilibrated'' on a lattice of size $L\times L$ with an MCMC step number of $10L^2$. Thus, it is hoped that there are no observable correlations between configurations in the chosen small fraction of replicas. Indeed, our results support this assumption. 

Note that one can generate samples using the multicanonical algorithm~\cite{Berg-1992,Berg-1993}.  Other algorithms~\cite{Landau-Binder-book}, including the Wang-Landau algorithm~\cite{WL-1,WL-2}, can also be used. In this case, it is necessary to provide a protocol for selecting uncorrelated samples. For example, in the Wang-Landau algorithm, which is a random walker in energy space, visiting the next energy level results in a local snapshot change, and there is a finite probability of returning to the first level with a different but still local snapshot change, so strong correlations obviously exist. In the case of MCPA, replicas at a given energy are weakly correlated, as are replicas at subsequent energy levels~\cite{Rose-2019,MS-2024,Fadeeva-2024}.

\section{Potts model} 

We consider the Potts model~\cite{Potts-1952} on a square lattice $L\times L$ with periodic boundaries; spins $s_i \in \{0,\dots,q{-}1\}$; summation over all pairs of spins $(s_i,s_j)$ with Hamiltonian $H=-\sum_{<i,j>}\delta_{s_i,s_j}$. The model undergoes a first-order phase transition when the number of spin components $q\ge 5$~\cite{Baxter-book}.  The value of the latent heat is small at $q=5$, and the correlation length depends significantly on $q$. The magnitude of the energy jump from the energy of the ordered phase $e_o$ to the energy of the disordered phase $e_d$ increases as the number of components $q$ increases~\cite{Baxter-book}
\begin{equation}
    e_d - e_o = 2\left( 1+\frac{1}{\sqrt{q}}\right)\tanh\left( \frac{\Theta}{2}\right) \prod^{\infty}_{n=1}{\tanh^2(n\Theta)},
    \label{eq:e_delta}
\end{equation}
where $\Theta$ is defined as $2\cosh(\Theta) {=} \sqrt{q}$.  Using the additional exact expression
\begin{equation}
    \frac{e_o+e_d}2=-\left( 1+\frac{1}{\sqrt{q}}\right)
    \label{eq:e_mean}
\end{equation}
we obtain the numerical values of $e_o$ and $e_d$ for  $q=10$ and $q=20$, which will be used to train the neural network model with ternary classification. We give the corresponding values in Table~\ref{table1}. In the following, we will refer to these models as PM-10 and PM-20.

The correlation length in the case of the first-order phase transition is finite, and for the Potts model it is decreases with increasing number of spin components. The last column in Table~\ref{table1} illustrates this behaviour with numerical values of the correlation length $\xi$ calculated  using the analytical expressions obtained in the paper~\cite{BJ-1992}. 

These values demonstrate our choice of the PM-10 and PM-20 models, which have a relatively small value for the correlation length, which must be smaller than the size of the lattices available in the simulation. This size is limited by the available memory of the computing resources. On the other hand, the larger the number of spin components $q$ is, the larger the computation time at each elementary step. Thus, our choice of $q$ and $L$ is a primitive optimization of the simulation time. 

\begin{table}[]
\caption{ Exact values of ordered energy, disordered energy, latent heat, and correlation length $\xi$~\cite{BJ-1992} for $q$-state Potts model. }
\begin{tabular}{|r||r|r|r||r|}
\hline
q  & $e_o$ & $e_d$ & $e_d-e_o$ & $\xi$ \\ \hline
5 & -1.473673 & -1.420754 & 0.052919 & 2512.2 \\ \hline
6 & -1.508980 & -1.307516 & 0.201464 & 158.9 \\ \hline
8 & -1.596732 & -1.110374 & 0.486358 & 23.9 \\ \hline
10 & -1.664252 & -0.968203 & 0.696049 & 10.6 \\ \hline
12 & -1.713644 & -0.863706 & 0.849938 & 6.5\\ \hline
13 & -1.733428 & -0.821272 & 0.912156 & 5.5 \\ \hline
20 & -1.820684 & -0.626529 & 1.194155 & 2.7 \\ \hline
\end{tabular}
\label{table1}
\end{table}

\section{Supervised machine learning of discontinuous phase transitions}

The main difference between phase transitions of the first order and phase transitions of the second order is in the completely different behavior of the correlation length. In the case of phase transitions of the second order, due to the tendency of the correlation length to infinity at the critical point, there is no physical scale at this point. Mathematically, this leads to the fact that the free energy function is described by a scaling form. A consequence of this mathematical structure is the classification of physical systems and models into universality classes. Thus, the most interesting physical studies of such systems are in the vicinity of the critical temperature, which is traditionally studied experimentally and theoretically, as well as by numerical modeling and machine learning.

In the case of phase transitions of the first  order, the critical behavior is quite different. Because of the finiteness of the correlation length, the free energy function does not scale with nontrivial critical exponents, and the notion of universality is still unknown, if it exists at all. The main feature of the critical point is the jump in internal energy, and the magnitude of the energy jump (equal to the latent heat) characterizes the strength of the transition: the larger the jump, the stronger the phase transition. Below it will be demonstrated on the example of the Potts model, in which the correlation length and the latent heat depend on the number of spin components.

Thus, the first-order phase transition is characterized by the value of the critical point and the magnitude of the internal energy jump. The temperature dependence of the free energy is the same for any physical system and model. Therefore, it is logical to look at the phase transition from a different angle. Instead of temperature dependence of physical quantities, we study the behavior of the system with energy dependence~\cite{Rose-2019}.

In the case of the first order phase transition, we have three phases separated by two energies~\cite{LL5}. Therefore, it is natural to use a ternary classification of samples, using an ensemble of physical model snapshots in the energy space, and to estimate the probabilities of the three phases as functions of energy.

\subsection{Ternary classification} 
The methodology of applying supervised machine learning to study phase transitions of the first  order consists of the following steps: 
\begin{enumerate}  
\item generates a large number of snapshots of spin configurations (samples) of the lattice model with energy $e$;
\item divides the samples into datasets for training/learning the neural network model and datasets for testing;
\item train the neural network model for ternary classification of the training dataset;
\item testing the training dataset to obtain predictions of the phases;
\item calculating the probability of belonging to the ordered, coexistence and disordered phases. The second moment of the predictions is also calculated;
\item obtaining numerical values of energy of ordered and disordered phases.
\end{enumerate}

\subsection{Sample generation} 

The MCPA method~\cite{Rose-2019,MS-2024} is used to obtain samples. Details and performance of the algorithm as applied to the Potts model  with 10 and 20 states can be found in the paper~\cite{Fadeeva-2024}.   For the present study, it is important to know that our implementation of data generation allows us to obtain a large number of samples at a given energy, and this rather large dataset represents almost uncorrelated samples. This is a consequence of the fact that we model a large number of replicates in the population and use a Monte Carlo step randomization process that formally occurs at infinite temperature. 

\begin{figure}
\includegraphics[width=.7\linewidth]{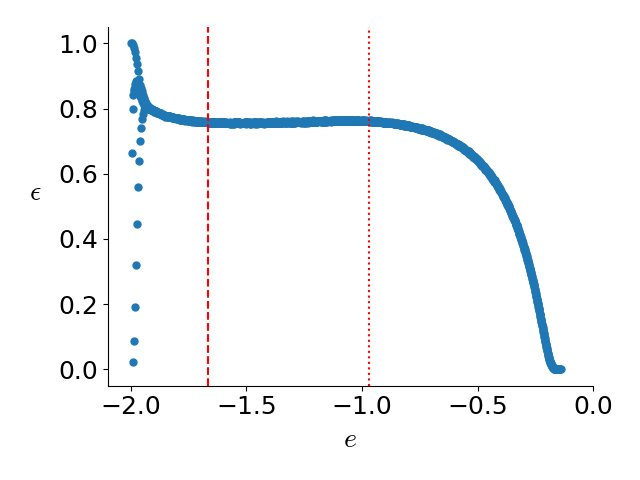}
\includegraphics[width=.7\linewidth]{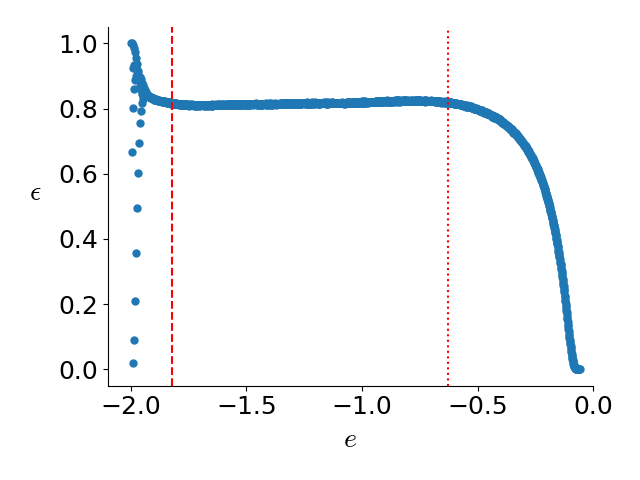}
\caption{ The fraction of $\epsilon$ configurations in the population with energy $e$. The red vertical lines indicate the critical energies $e_o$ and $e_d$. Top figure: PM-10 model, bottom figure: PM-20 model.}
\label{fig1}
\end{figure}

At each energy value, we randomly select $N=8192$ snapshots from the total number of configurations at energy $e$. The Figure~\ref{fig1} shows the so-called cooling factor $\epsilon$~\cite{Rose-2019,MS-2024}, which in our case is appropriately called the fraction of configurations at energy $e$. In the energy region of interest $-2<e<-0.5$ this fraction is about $\epsilon\approx 0.8$, then in a total of $2^{17}$ replicas in MCPA simulations about $10^5$ replicas have energy $e$ of interest. From these, we select about every tenth replica to form a dataset. This data set is divided in a 3:1 ratio, i.e., $N_l=6144$ spin configurations are used for training (learning) and $N_p=2048$ configurations are used for predictions and probability estimation. 

 \begin{figure}
\center
\includegraphics[width=1\linewidth]{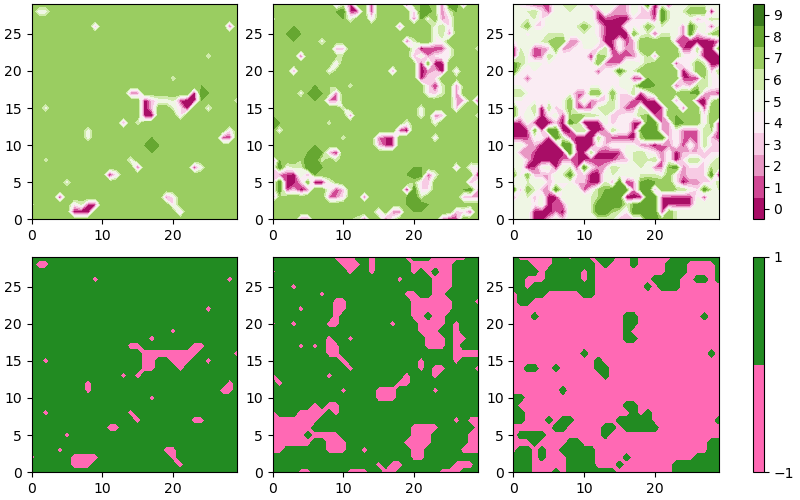}
\caption{Typical spin configurations for the 10-state Potts model on the $L=30$ lattice at energies from left to right: $e=-1.9$ in the ordered phase, $e=-1.4$ in the coexistence phase, and $e=-0.9$ in the disordered phase. The upper panel is the raw dataset RD and lower panel is the majority/minority MD dataset.  The vertical colored bar marks the spin number.}
\label{fig2}
\end{figure}
\begin{figure}
\center
\includegraphics[width=.7\linewidth]{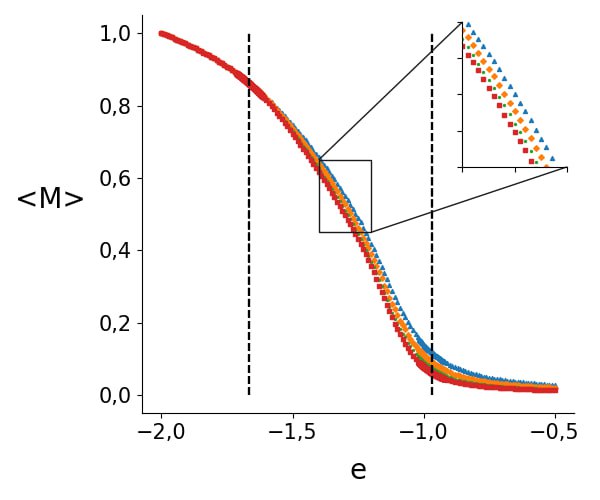}
\includegraphics[width=.7\linewidth]{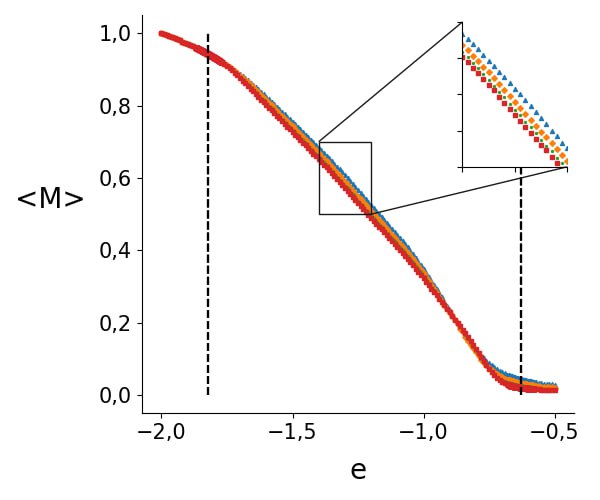}
\caption{ Average magnetization $<M>$ of the samples at a given energy $e$. The insets show symbols and colors for grid sizes $L=30$ - blue triangles, 40 - orange rhombuses, 50 - green dots and 60 - red squares. Top figure: model PM-10, bottom figure: model PM-20.}
\label{fig3}
\end{figure}
 
\subsection{Data preprocessing} Typical PM-10 configurations corresponding to ordered, coexistence and disordered states are shown in the Figure~\ref{fig2}, the color scale on the right corresponds to the instantaneous spin values.
 There are two ways to represent spin configuration for Potts model, which we marked with the abbreviations: RD - the raw data and MD - in each configuration the spins belonging to the largest component $m$ in $q=0,1,\ldots q-1$ is marked as +1 and the rest of spins with the 
-1. The second line of Figure~\ref{fig2} illustrates the result of the top line with raw RD configurations mapped to majority/minority MD configurations in all three phases.

This representation is inspired by the way the magnetisation is calculated in simulations of the Potts model~\cite{Binder-1981} with spin $m$ majority by $M=(q N_m/L^2-1)/(q-1)$, where $N_m$ is the number of sites $i$ with $s_i=m$. The dependence of magnetization of average magnetization $<M(e)>=\sum_1^N M(e)/N$, calculated over all configurations with energy $e$, is plotted in Figure~\ref{fig3} for several values of $L$ and for two models, PM-10 and PM-20. Note that the majority choice is random,  and in Figure 2, the green color in the bottom panel reflects the majority's backs, which corresponds to $m=7$ in the left and middle figures of the top panel and $m=5$ in the right figure.

Note that the geometry of the two representations is quite different and yet, as will be seen below, training and testing on the two datasets leads to quantitatively same predictions of the model phases. At the same time, in the case of MD datasets, it is impossible to extract complete information from the sample from the bottom row of Figure~\ref{fig2}. For example, it is impossible to correctly calculate the total energy. 
In the scientific community of physicists, the question is being discussed: what does a neural network “see” (This jargon is appropriate, since neural networks were rapidly developed in the context of machine vision)? Our view, which we are working to confirm, is that a neural network sees correlations and typical dimensions of geometric objects. In the case of a phase transition of the second order, this results in the second moment of the probability of belonging to the ferromagnetic (paramagnetic phase) having a width that scales with the critical exponent of correlation length~\cite{Chertenkov-2023}. In the case of phase transitions of the first order, the correlations encoded in the rounding of the critical region, as was shown for the first time in Imry paper~\cite{Imry-1980}.  

\subsection{Training neural network model} 

 We train the neural network independently for each model PM-10 and PM-20 and for each lattice size $L=30,40,50$ and 60, and for two different datasets, RD and MD. Thus, we train a total of 16 NN models that can be annotated by NN-q-L-Dataset, with $q$ denotes the number of spin components, $L$ denotes the lattice size, and Dataset denotes RD or MD.    

Specifically, the four models, NN-10-30-RD, NN-10-40-RD, NN-10-50-RD, and NN-10-60-RD, use the datasets created for the PM-10 model in RD representation on lattices of sizes 30, 40, 50, and 60, respectively.
For the PM-20 model and RD dataset, we train four other NN models, NN-20-30-RD, NN-20-40-RD,  NN-20-50-RD, and NN-20-60-RD, by analogy. Correspondingly, for the MD dataset, we use abbreviations such as NN-10-30-MD, NN-10-40-MD, ..., NN-20-30-MD, ... NN-20-60-MD. At each energy value  $e$, each dataset consists of $N_l=6144$ samples. 

We use CNN neural network~\cite{CNN} with binary cross-entropy loss function and Adam optimization algorithm~\cite{Adam-2014} with parameters $\alpha{=}10^{-3}$, $\beta_1{=}0.9$, $\beta_2{=}0.999$, and $\varepsilon{=}10^{-8}$.  The network is trained in one epoch~\cite{epoch}. 

The samples in each training dataset are labeled as belonging to one of three phases - ordered, coexisting, and disordered, abbreviated OS, CS, and DS, respectively. To classify the images into the three phases, we use the precisely known energy values $e_o$ and $e_d$ given in Table~\ref{table1}. The image simulated at energy $e$, labeled as
\begin{itemize}
\item[OS] if $e < e_o$;
\item[CS] if $e_o<e < e_d$;
\item[DS] if $e> e_d$.
\end{itemize}
  It should be noted that the energy value is not transferred to the NN during training. When training a neural network model, only the label of belonging to one of the three phases is transmitted. 

\section{Probabilities of the ordered, coexistence, and disordered states} 

In this section, we propose a way to estimate the probabilities of ordered, coexistence, and disordered phases of matter as functions of energy. To do this, we use the predictions of a neural network model about whether a particular sample belongs to one of the three phases. Taking an ensemble of samples at energy $e$, we estimate the probability that this ensemble belongs to one of the three phases. Varying the energy, we obtain probabilities as functions of energy. 

In addition, we estimate variation of the predictions. From the probability function and variation function, we extract the  critical energy of the disordered state and itical energy of the ordered state, thus estimating the value of the latent heat. The variation function gives us clear evidence of thermally induced fluctuations in the vicinity of this two energies~\cite{LL5,Imry-1980}. This functions can be fitted to a Gaussian, which gives us an estimate of the width of thermal fluctuations, which are inversely proportional to latent heat.

\subsection{Predictions and phase probabilities}

 It was stated above that 75\% of the samples were used for training the neural network. The remaining 25\% of the samples are used for predictions. Each configuration $i$ of size $L\times L$ sampled in MCPA simulations with energy $e$ is fed to the input of a trained NN(L) network. The network outputs three numbers predicting membership of the ordered phase $p_{OS}^i(e)$, the coexistence phase $p_{CS}^i(e)$, and the disordered phase $p_{DS}^i(e)$.  The sum of these three numbers is equal to one. We repeat this process over a sample space containing $N_{test}=2048$ samples with a given energy $e$. 

\begin{figure}
\center
\includegraphics[width=0.5\linewidth]{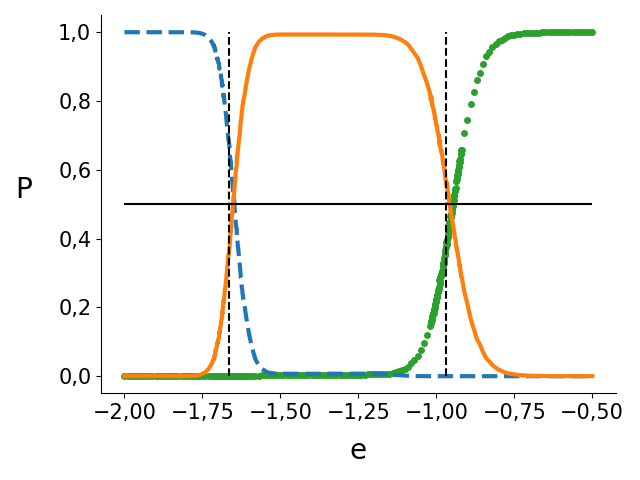}~\includegraphics[width=.5\linewidth]{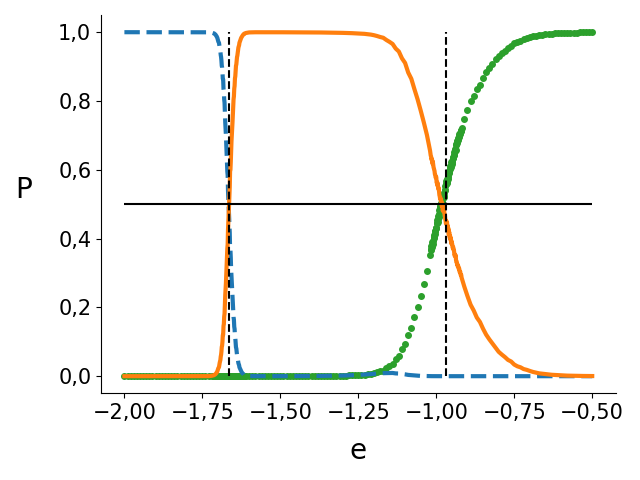}
\includegraphics[width=0.5\linewidth]{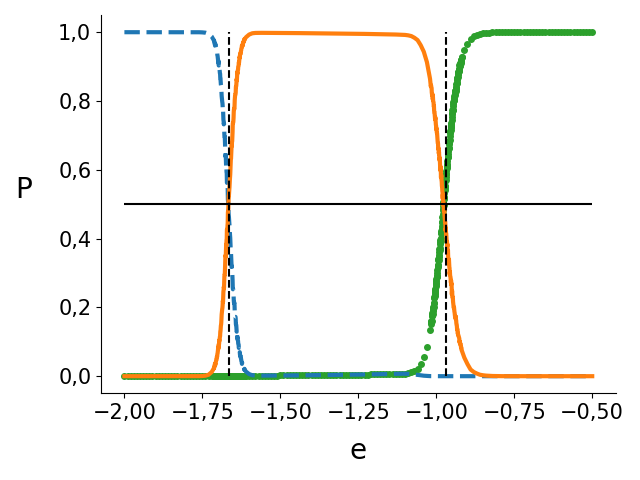}~\includegraphics[width=.5\linewidth]{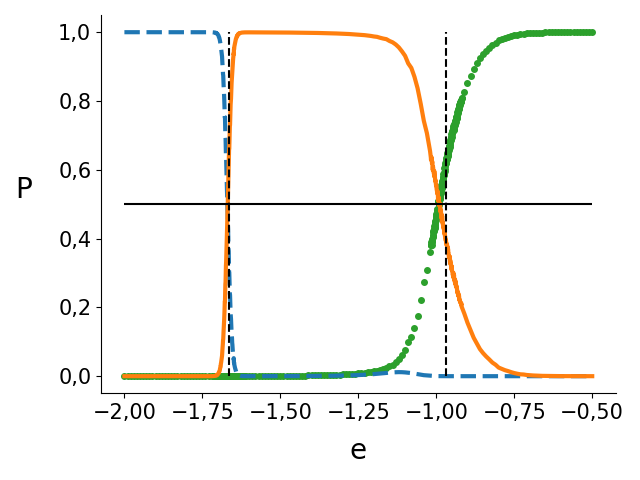}
\includegraphics[width=0.5\linewidth]{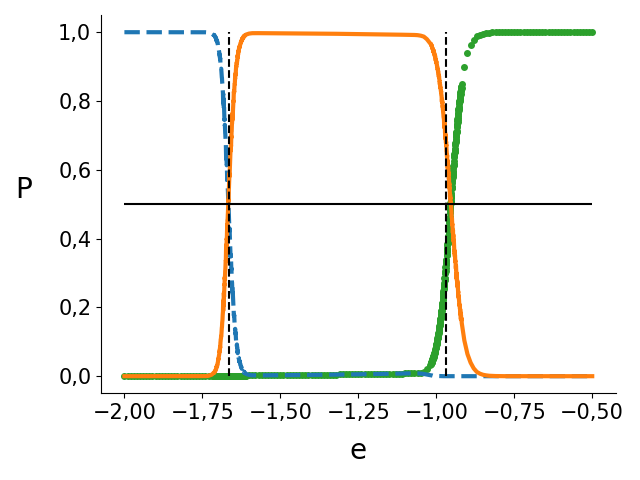}~\includegraphics[width=.5\linewidth]{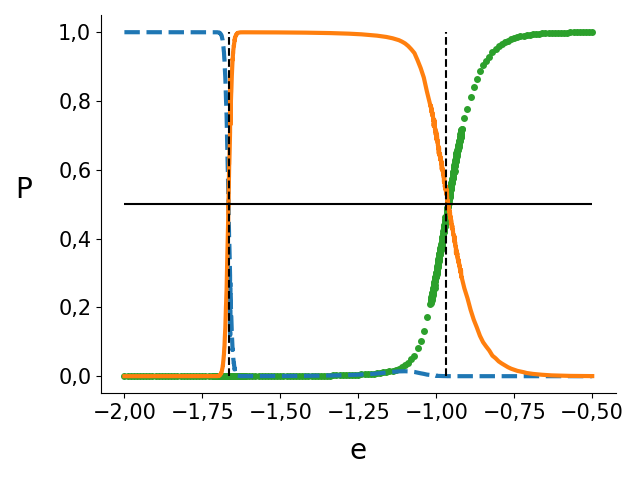}
\includegraphics[width=0.5\linewidth]{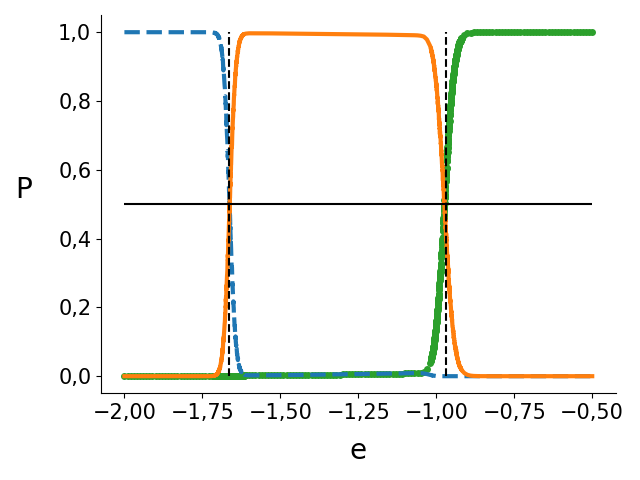}~\includegraphics[width=.5\linewidth]{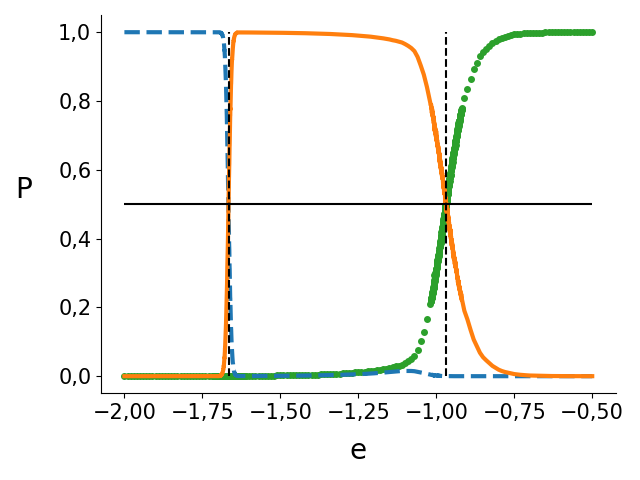}
\caption{Probabilities of phases $P_{xS}(E)$ for $L=30$ (first row), 40 (second row), 50 (third row) and 60 (last row) for 10-state Potts model, PM-10. Left panel is the training/testing with the raw dataset RD and right panel is the training/testing with the majority/minority dataset MD.  Blue dashed lines are ordered phase probabilities $P_{OS}$, orange solid lines are coexistence phase probabilities $P_{CS}$, and green dotted lines are disordered phase probabilities $P_{DS}$. The vertical lines denote the exact values of the critical energies of the ordered and disordered phase.}
\label{fig4}
\end{figure}

\begin{figure}
\center
\includegraphics[width=.5\linewidth]{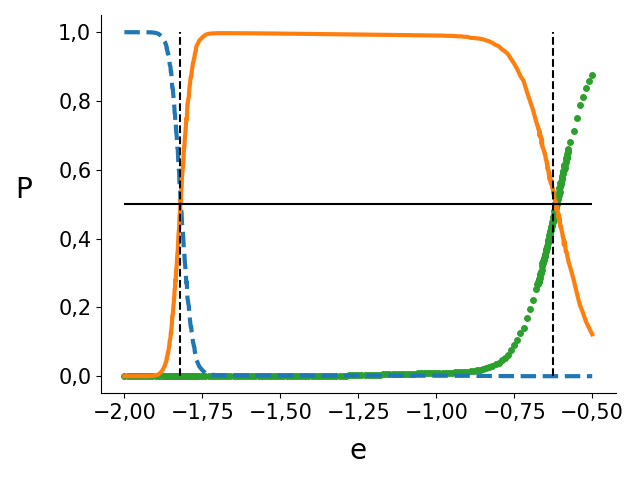}~\includegraphics[width=.5\linewidth]{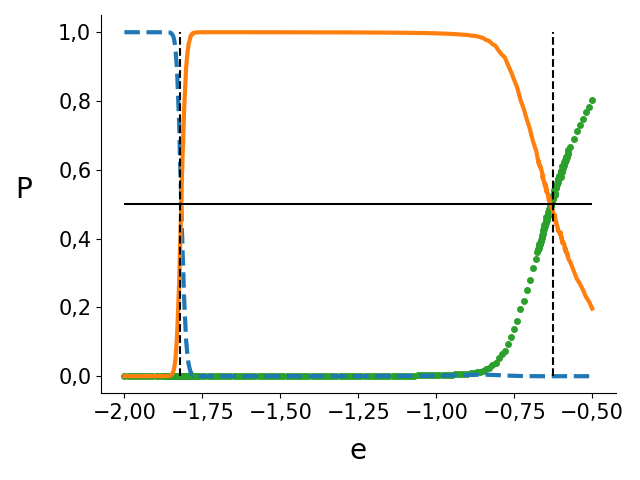}
\includegraphics[width=.5\linewidth]{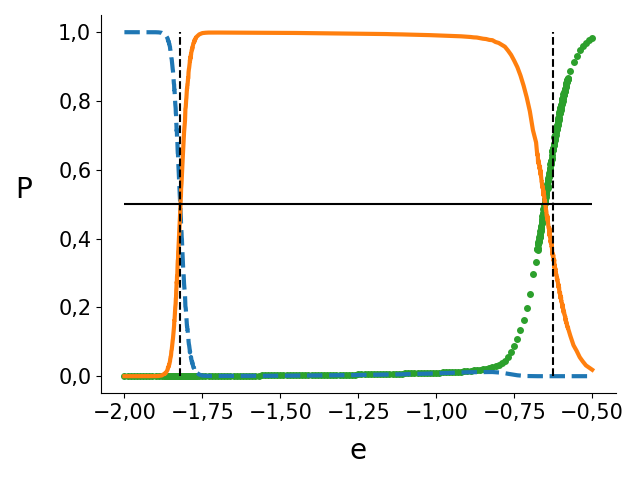}~\includegraphics[width=.5\linewidth]{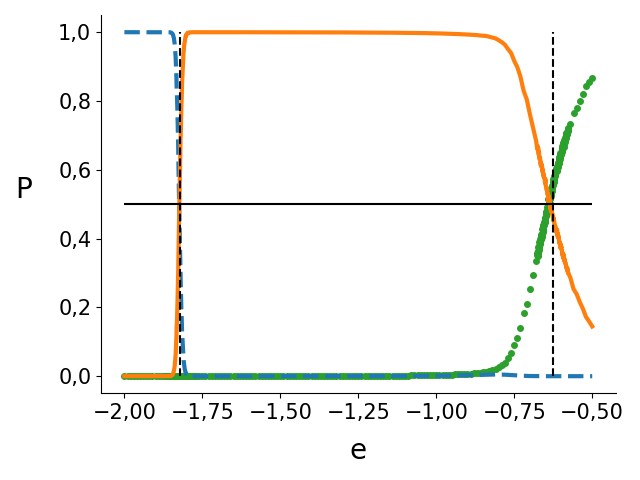}
\includegraphics[width=.5\linewidth]{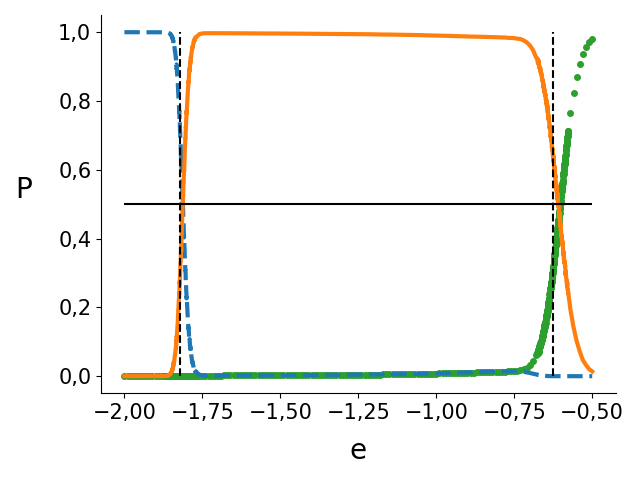}~\includegraphics[width=.5\linewidth]{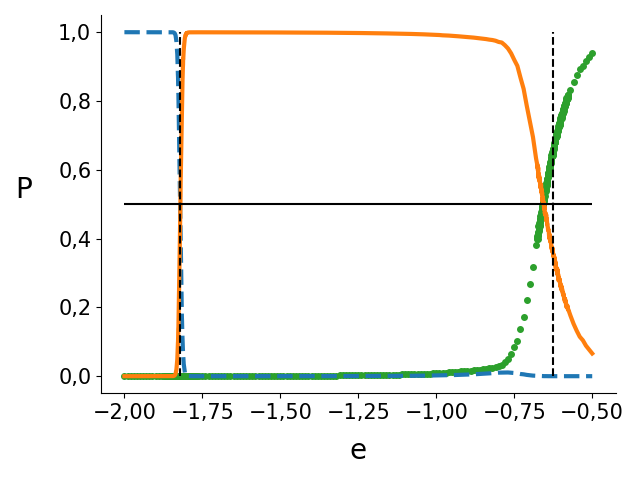}
\includegraphics[width=.5\linewidth]{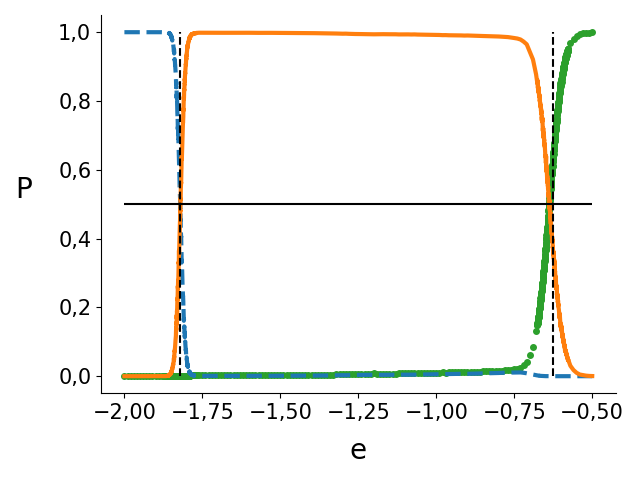}~\includegraphics[width=.5\linewidth]{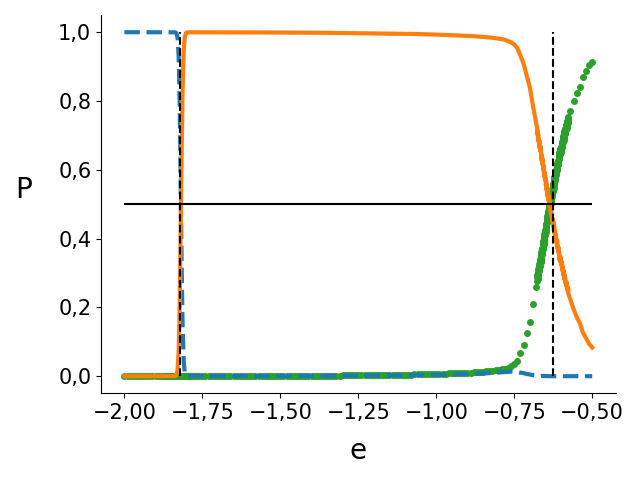}
\caption{Same as in Fig.~\ref{fig4}  for 20-state Potts model, PM-20.}
\label{fig5}
\end{figure}

We calculated estimates of the probabilities $P_{xS}(e)$ averaging predictions $p_{xS}^i(e)$ over the sample space, 
\begin{equation}
P_{xS}(e)=\frac{1}{N_{test}}\sum_{i=1}^{N_{test}}p_{xS}^i(e),
\label{eq:prob}
\end{equation} 
where $xS$ stands for one of the three phases, $OS, CS$, or $DS$. 
Thus, we obtain the probability functions $P_{OS}(e)$, $P_{CS}(e)$, and $P_{DS}(e)$, with $P_{OS}(e)+P_{CS}(e)+P_{DS}(e)=1$. Figures \ref{fig4} and \ref{fig5} show an example of phase probability estimation for model PM-10 and model PM-20, respectively, and present the results for both datasets, RD and MD. 
The network correctly predicts the phases of the models, and the smoothed sharp change of probabilities near critical energies is explained by the finiteness of the system under study.  Note that, to the best of our knowledge, phase probabilities have not been studied before.

\subsection{Energy and latent heat estimations} To estimate the energy of the ordered phase $e_o$, we chose several (in fact we stopped at seven) points belonging to each set $P_{OS}(e)$ and $P_{CS}(e)$ in the neighborhood of their intersection, and approximated these points by two straight lines, respectively. The intersection of these straight lines gives an estimate of $e_o$. Similarly, the estimate for $e_d$ is obtained from the sets $P_{CS}(e)$ and $P_{DS}(e)$. The resulting estimates are presented in the tables.

Tables \ref{table2} and \ref{table3} present estimates of critical energies $e_o$ and $e_d$ and latent heat for the 10-component Potts model, PM-10. Each row of critical energy estimates is followed by a row with the ratio of the deviation of the estimated critical energy to the statistical error to show the quality of the estimate.  The estimates are scattered around a precisely known value of the critical energy. We found no noticeable finite-size dependence. 

In the case of a continuous transition driven by thermodynamic fluctuations, we see that finite-size analysis of machine learning probabilities reflects these fluctuations and leads to a reasonable estimate of the critical length exponent~\cite{Carrasquilla-2017,Chertenkov-2023}. In the case of a discontinuous phase transition, the correlation length does not diverges, and the dependence on the finite size may be different. It is known that finite-size corrections to thermodynamic quantities in the Potts model are very sensitive to the way they are estimated and depend on the quantities of interest. In two-dimensional space, the corrections can be proportional to $1/L$ or $1/L^2$ or even $1/L^4$~\cite{Challa-1986,Lee-1991,BJ-1992,Fadeeva-2024}. A possible explanation for the apparent lack of finite size dependence of probability is that, for some reason, the amplitude $A$ of the correction to the probability function, which, by analogy  with the energy  probability density function,  should behave as $A/L$~\cite{Lee-1991}, is small. Nevertheless, we will see below that thermal fluctuations do depend on the lattice size, as can be seen from the behaviour of the variation $D_{xS}(E)$ of predictions. 

 The phase prediction probability of a neural network is not a thermodynamic function, but at the same time its fluctuations can somehow reflect thermodynamic fluctuations, which may lead to the visible finite-size corrections. The correlation length is finite and approximately equal to 10.6 for the PM-10 model~\cite{BJ-1992}, see also Table~\ref{table1}. Interestingly, even moderate sizes of the studied systems compared to the correlation length allow us to estimate the critical values of the energies, and through them the latent heat, with an accuracy not worse than few percent. 

\begin{table}[h!]
\center
\begin{tabular}{|c||cccc|c|} 
\hline 
$L$    & 30& 40 & 50 & 60 & Exact \\ \hline \hline
$e_o$ & -1.650(20) & -1.667(6) & -1.667(6) & -1.663(6) & -1.66425\ldots \\ \hline
$\frac{\Delta e_o}{\sigma_o}$ & 0.7 & 0.5 & 0.5 & 0.2 & \\ \hline
$e_d$ & -0.952(28) & -0.977(1) & -0.954(5) & -0.974(1) & -0.96820\ldots \\ \hline
$\frac{\Delta e_d}{\sigma_d}$ & 0.6 & 8.8 & 2.8 & 5.8 & \\ \hline
$\cal L$ & 0.698(48) & 0.690(7) & 0.713(11) & 0.689(7) & 0.696049\ldots\\ \hline
$\Delta \cal L$ & 0.0002 & -0.006 &   0.017 & -0.007 & \\ \hline
 \end{tabular}
\caption{Estimates of the critical energies and latent heat for PM-10 using raw data set RD.}
\label{table2}
\end{table}

\begin{table}[h!]
\center
\begin{tabular}{|c||cccc|c|} 
\hline 
$L$    & 30& 40 & 50 & 60 & Exact \\ \hline \hline
$e_o$ & -1.665(17) & -1.669(21) & -1.666(3) & -1.666(7) & -1.66425\ldots \\ \hline
$\frac{\Delta e_o}{\sigma_o}$ & 0.0 & 0.2 & 0.6 & 0.2 & \\ \hline
$e_d$ & -0.983(5) & -0.992(9) & -0.962(2) & -0.969(3) & -0.96820\ldots \\ \hline
$\frac{\Delta e_d}{\sigma_d}$ & 3.0 & 2.6 & 3.1 & 0.3 & \\ \hline
$\cal L$ & 0.682(22) & 0.677(30) & 0.704(5) & 0.697(10) & 0.696049\ldots\\ \hline
$\Delta \cal L$ & -0.014 & -0.019 &   0.008 & 0.001 & \\ \hline
 \end{tabular}
\caption{Estimates of the critical energies and latent heat for PM-10 using majority/minority data set MD.}
\label{table3}
\end{table}

Tables \ref{table4} and \ref{table5} present estimates of critical energies $e_o$ and $e_d$ and latent heat for the 20-component Potts model, PM-20.
The quality of estimates is as good as those in the case of the PM-10 model. Again, our data show no noticeable dependence of the estimates on the system size. In this case, the correlation length in the critical point is much smaller at about 2.7~\cite{BJ-1992}, and yet we see no finite-size corrections in our estimates. In contrast, in our recent study on PM-10 and PM-20~\cite{Fadeeva-2024} using the Wang-Landau algorithm in simulations and the same MCPA algorithm as in the present research, the finite-size corrections to the critical energies estimated from the energy probability distribution show $1/L$ corrections, as predicted by the analytics~\cite{Challa-1986,Lee-1991}.

\begin{table}[h!]
\center
\begin{tabular}{|c||cccc|c|} 
\hline 
$L$    & 30& 40 & 50 & 60 & Exact \\ \hline \hline
$e_o$ & -1.820(12) & -1.821(11) & -1.813(2) & -1.821(15) & -1.82068\ldots \\ \hline
$\frac{\Delta e_o}{\sigma_o}$ & 0.1 & 0.0 & 3.8 & 0.0 & \\ \hline
$e_d$ & -0.616(3) & -0.652(6) & -0.606(26) & -0.638(3) & -0.626529\ldots \\ \hline
$\frac{\Delta e_d}{\sigma_d}$ & 3.5 & 4.2 & 0.8 & 3.8 & \\ \hline
$\cal L$ & 1.204(15) & 1.169(17) & 1.207(28) & 1.183(18) & 1.19415\ldots\\ \hline
$\Delta \cal L$ & 0.010 & -0.025&   0.013 & -0.011 & \\ \hline
 \end{tabular}
\caption{Estimates of the critical energies and latent heat for PM-20 using raw data set RD.}
\label{table4}
\end{table}

\begin{table}[h!]
\center
\begin{tabular}{|c||cccc|c|} 
\hline 
$L$    & 30& 40 & 50 & 60 & Exact \\ \hline \hline
$e_o$ & -1.818(2) & -1.824(17) & -1.821(31) & -1.819(8) & -1.82068\ldots \\ \hline
$\frac{\Delta e_o}{\sigma_o}$ & 1.3 & 0.2 & 0.0 & 0.2 & \\ \hline
$e_d$ & -0.632(1) & -0.638(12) & -0.657(7) & -0.636(3) & -0.626529\ldots \\ \hline
$\frac{\Delta e_d}{\sigma_d}$ & 5.5 & 1.0 & 4.4 & 3.2 & \\ \hline
$\cal L$ & 1.186(3) & 1.186(29) & 1.164(38) & 1.183(11) & 1.19415\ldots\\ \hline
$\Delta \cal L$ & -0.008 & -0.008&   -0.030 & -0.011 & \\ \hline
 \end{tabular}
\caption{Estimates of the critical energies and latent heat for PM-20 majority/minority data set MD.}
\label{table5}
\end{table}

\subsection{Finite-size rounding of the variation of the phase predictions}

In Imry's paper, it was shown~\cite{Imry-1980} that the rounding off of some thermodynamic functions in the case of the first-order phase transition is inversely proportional to the square of the system size times  the latent heat. Lee and Kosterlitz~\cite{Lee-1991} (see also~\cite{Challa-1986}) calculated this rounding in the more detail for the case of the Potts model, and, for example, the rounding of the specific heat is proportional to the logarithm of the number of states, divided by the latent heat and 
\begin{equation}
\Delta T \approx \frac{T_c \ln{q}}{{ \cal L} L^2} ,
\label{eq:LK}
\end{equation} 
and amaizingly,  the difference with Imry's formula is only in the multiplier with the number of components, which of course was outside Imry's 
analysis, which considered the general case of the first-order phase transition due to the thermal fluctuations and Landau theory~\cite{LL5}.

Analysis of the rounding in other quantities, such as Binder cumulants of energy and magnetization in the Potts model~\cite{Lee-1991}, shows that  the terms in the expression~(\ref{eq:LK}) remain for all thermodynamic quantities, and the expression can be modified by additional multipliers such as the  ratio of critical energies. Thus, it seems that the expression~(\ref{eq:LK})  is generally universal for the observables in the Potts model.

\begin{figure}
\center
\includegraphics[width=0.5\linewidth]{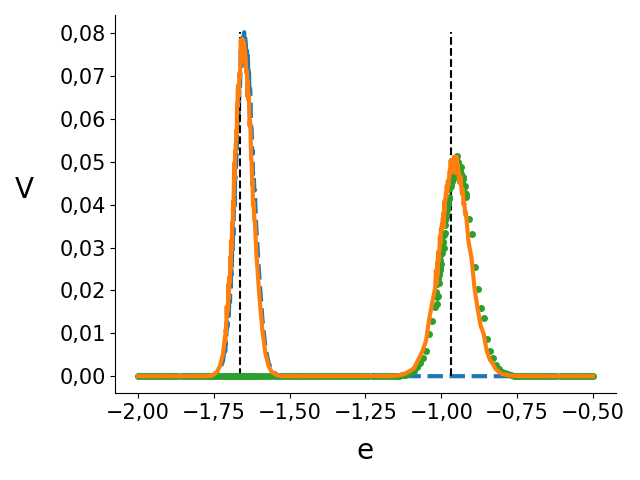}~\includegraphics[width=.5\linewidth]{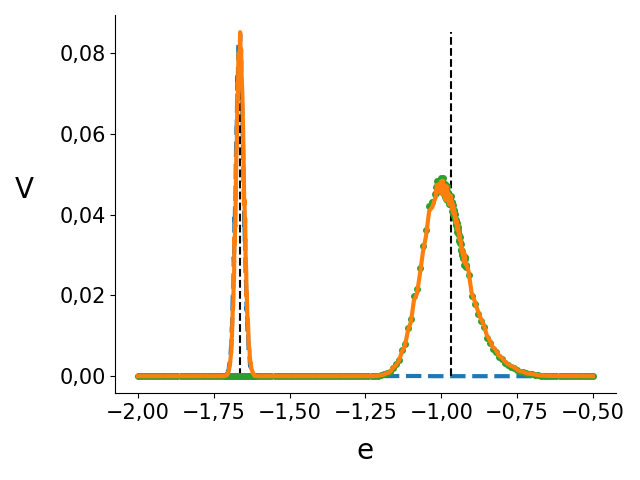}
\includegraphics[width=0.5\linewidth]{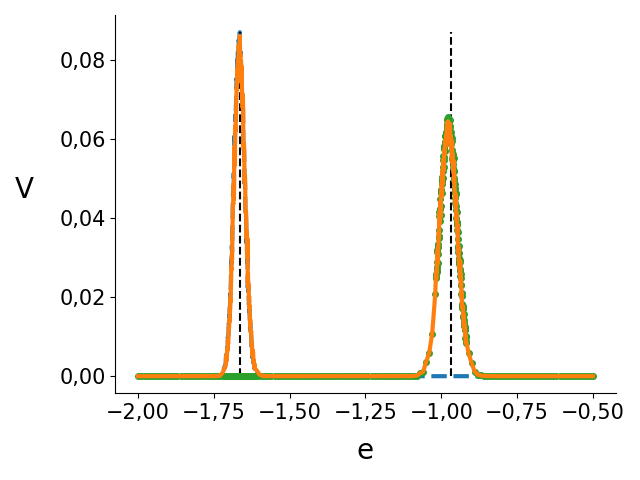}~\includegraphics[width=.5\linewidth]{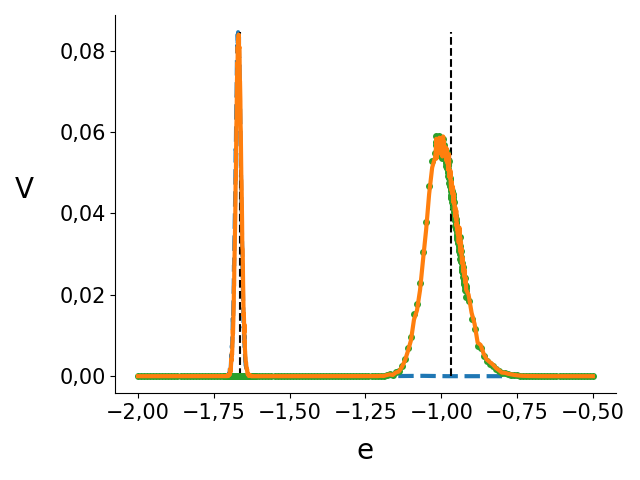}
\includegraphics[width=0.5\linewidth]{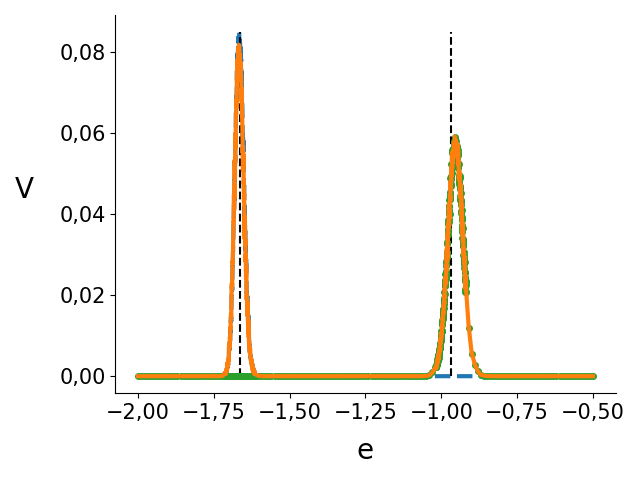}~\includegraphics[width=.5\linewidth]{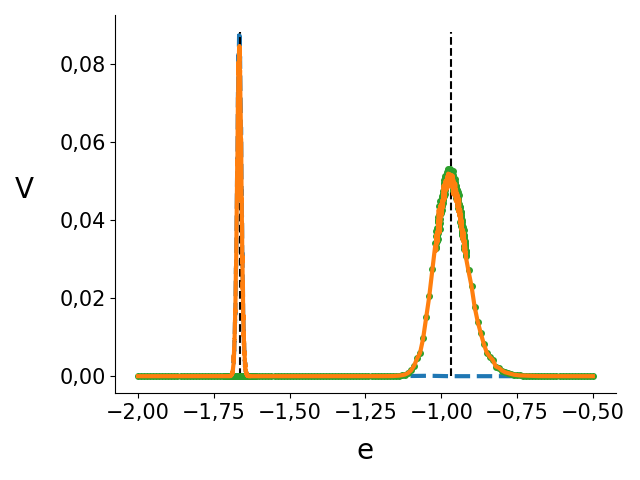}
\includegraphics[width=0.5\linewidth]{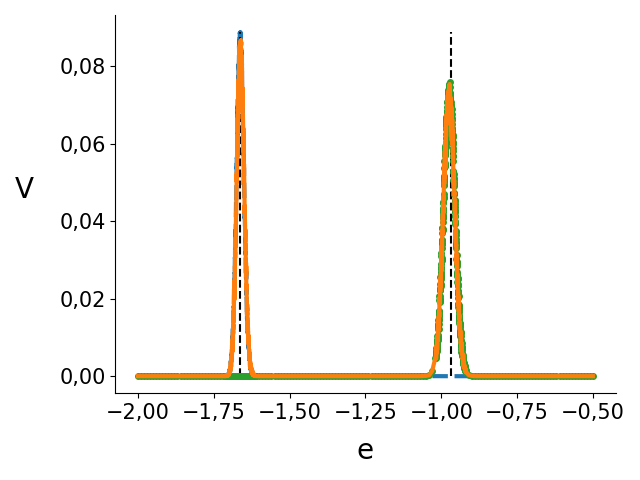}~\includegraphics[width=.5\linewidth]{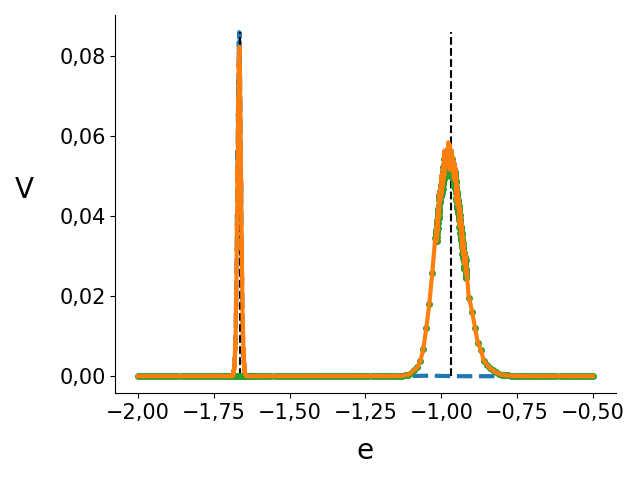}
\caption{ Variations $V_{xS}(E)$ of the probabilities of phases $P_{xS}(E)$ for $L=30$ (first row), 40 (second row), 50 (third row) and 60 (last row) for 10-state Potts model, PM-10. Left panel is the training/testing with the raw dataset RD and right panel is the training/testing with the majority/minority dataset MD.  Blue dashed lines are ordered  variations $P_{OS}$, orange solid lines are coexistence variations $P_{CS}$, and green dashed lines are disordered variations $P_{DS}$. The vertical lines denote the exact values of the critical energies of the ordered and disordered phase.}
\label{fig6}
\end{figure}

\begin{figure}
\center
\includegraphics[width=0.5\linewidth]{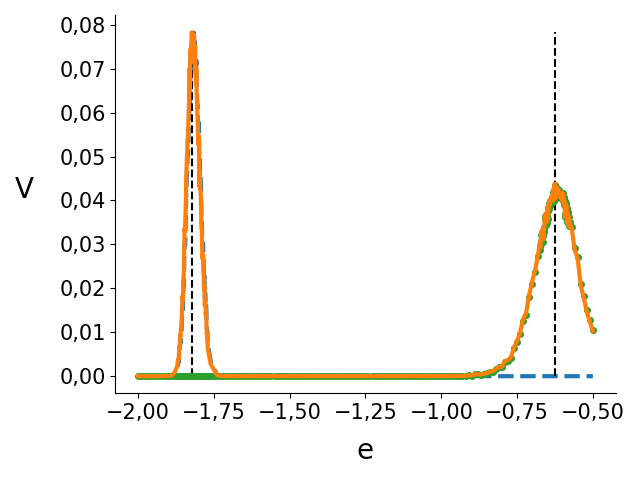}~\includegraphics[width=.5\linewidth]{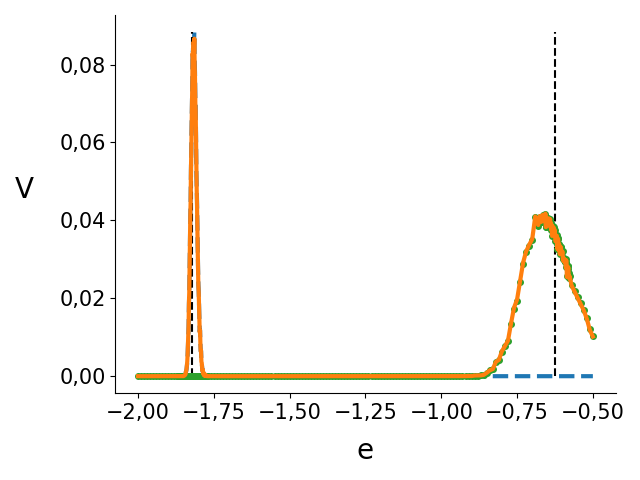}
\includegraphics[width=0.5\linewidth]{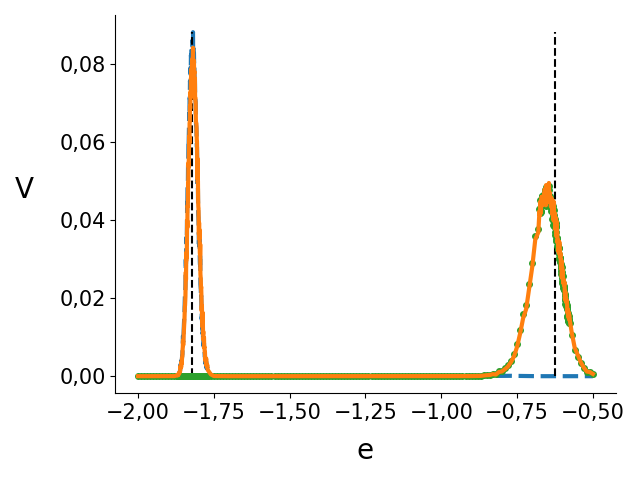}~\includegraphics[width=.5\linewidth]{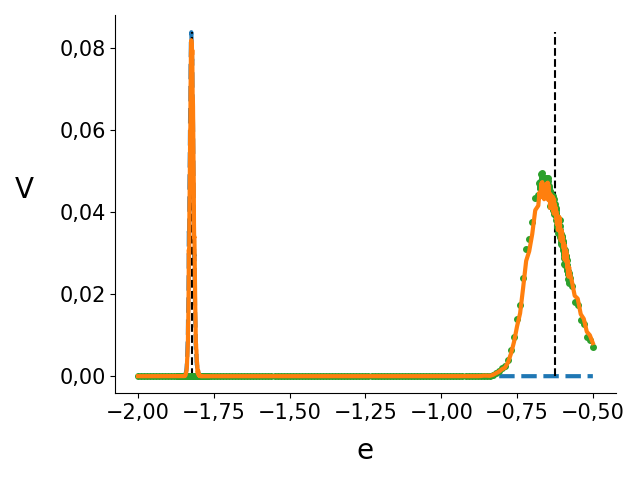}
\includegraphics[width=0.5\linewidth]{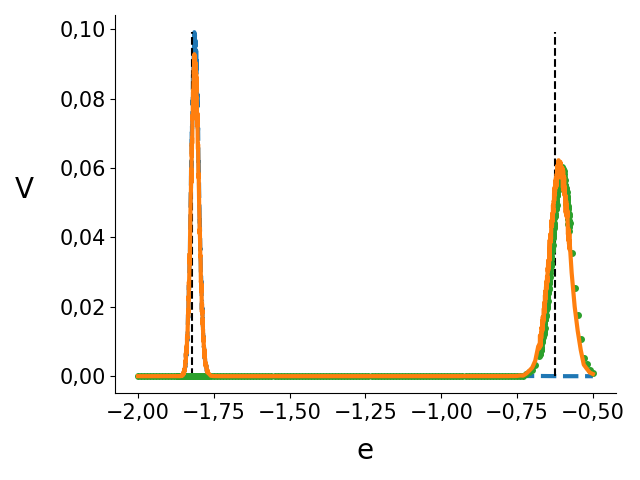}~\includegraphics[width=.5\linewidth]{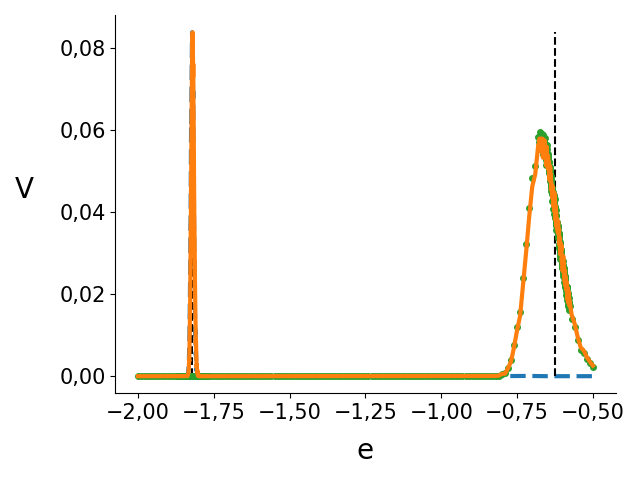}
\includegraphics[width=0.5\linewidth]{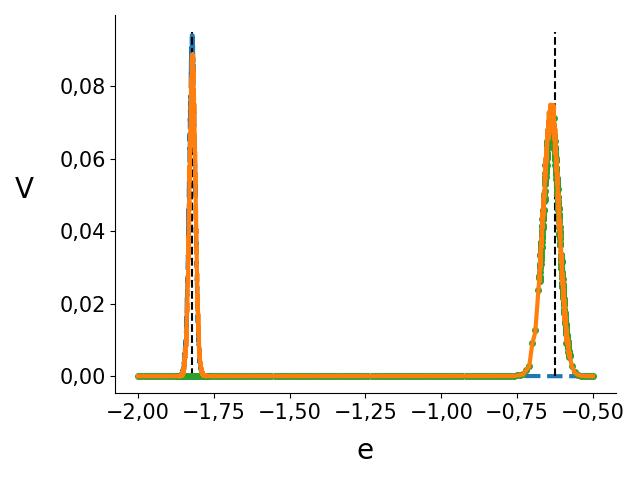}~\includegraphics[width=.5\linewidth]{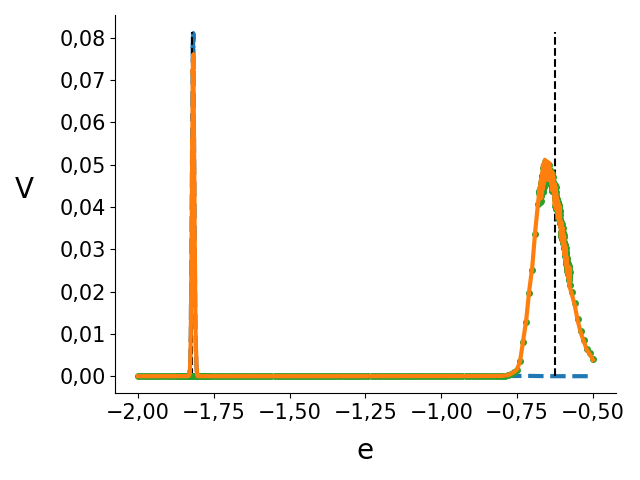}
\caption{Same as in Fig.~\ref{fig6}  for 20-state Potts model, PM-20.}
\label{fig7}
\end{figure}

As a result of the machine learning we estimates probabilities  $P_{xS}$ of the phases~(\ref{eq:prob}). Using parallel with the specific heat which is proportional to the variation of the energy, we can analyze variation $V_{xS}$ of the phase probabilities $P_{xS}$ (remind the reader that $xS$ stands for one of the three phases - OS, CS, and DS)

\begin{equation}
V_{xS}=\frac{1}{N_{test}}\sum_{i=1}^{N_{test}}\left[p_{xS}^i(e)\right]^2-\left[\frac{1}{N_{test}}\sum_{i=1}^{N_{test}}p_{xS}^i(e)\right]^2.
\label{eq:var}
\end{equation} 
 
Figures~\ref{fig6} and \ref{fig7} show the variation $V_{OS}$, $V_{CS}$, and $V_{DS}$ of the probabilities of the three phases as functions of energy for different lattice sizes for PM-10 and PM-20. The variation maximum occurs around the critical energies $e_d$ and $e_0$, with fluctuations around $e_d$ are more pronounced.  We use a Gaussian approximation in the neighborhood of the maximum, thus estimating the position of the maximum $\mu$ and its width $\sigma$. 
The position of the maximum $\mu$ gives estimates of the critical energies given in Tables~\ref{table6}-\ref{table9}, which are compatible with the estimates from the probabilities and given in Tables~\ref{table2}-\ref{table5}. 

Figures~~\ref{fig6} and \ref{fig7} show that the peaks become narrower as the lattice size increases. Indeed, the corresponding $\sigma$ values given in Tables~\ref{table6}-\ref{table9} show decreasing  peak widths. From these values we can get an idea of how the width depends on the latent heat $\cal L$, the number of spin components $q$ and the lattice size $L$.

\begin{figure}
\center
\includegraphics[width=0.5\linewidth]{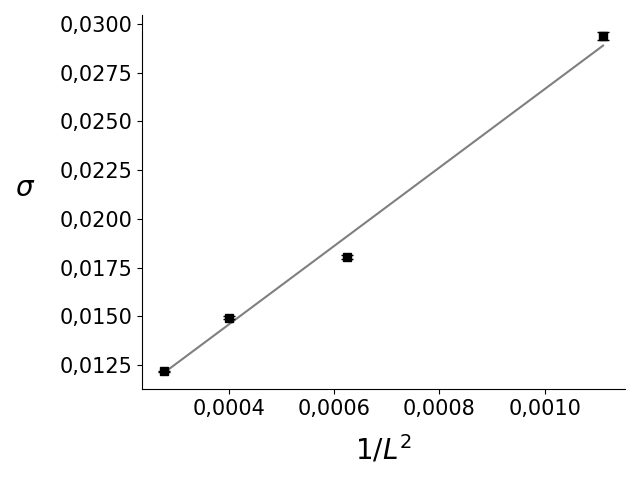}~\includegraphics[width=.5\linewidth]{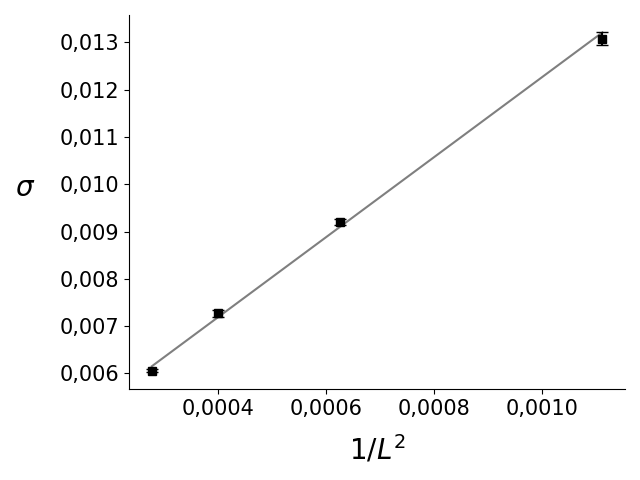}
\includegraphics[width=0.5\linewidth]{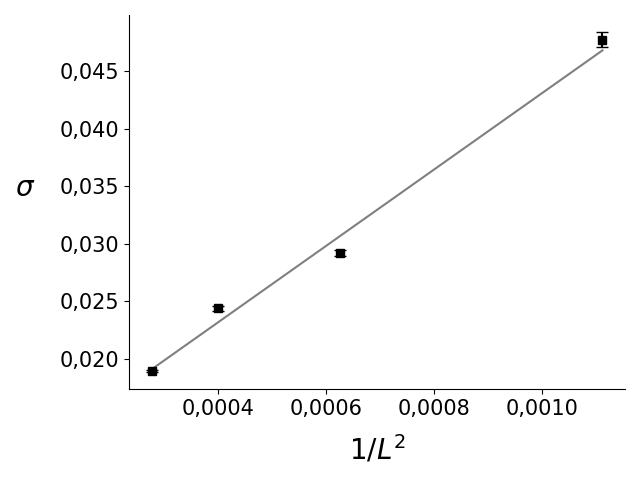}~\includegraphics[width=.5\linewidth]{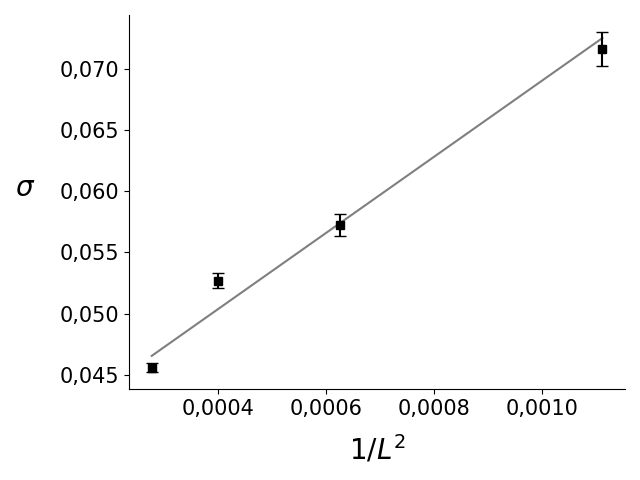}
\caption{Width $\sigma_{10}(L)$ of the variation $V_{CS}$ with error bars from Tables~\ref{table6} and \ref{table7} with the linear fit (solid line). Upper row represent data for the left peak in Figures~\ref{fig6} and bottom row represent data for the right peak in Figures~\ref{fig6}. Left row is the RD data set and right row is the MD data set. Data sets of PM-10 model, $q=10$. }
\label{fig8}
\end{figure}

\begin{figure}
\center
\includegraphics[width=0.5\linewidth]{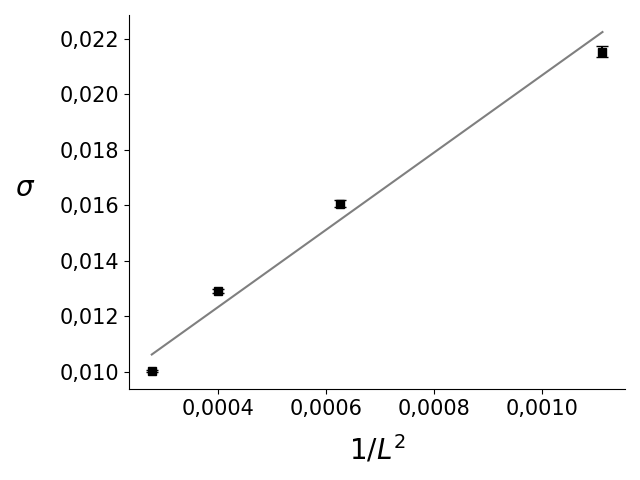}~\includegraphics[width=.5\linewidth]{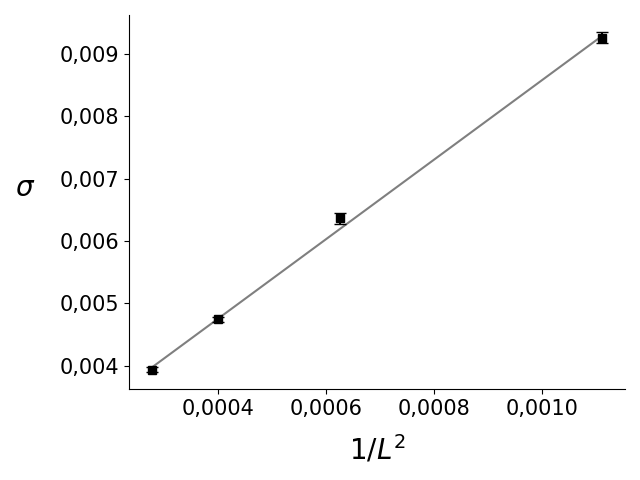}
\includegraphics[width=0.5\linewidth]{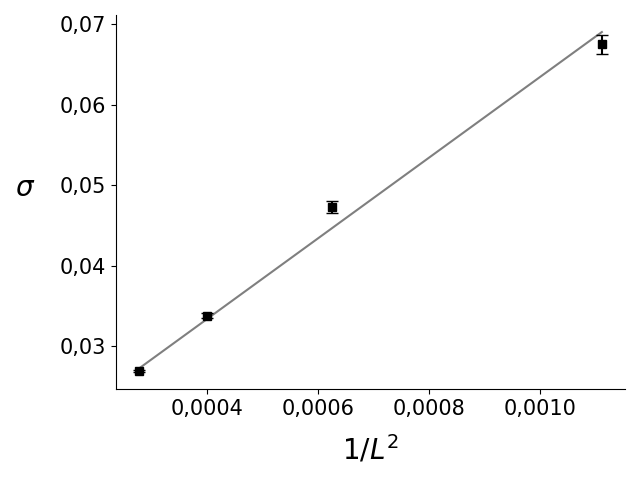}~\includegraphics[width=.5\linewidth]{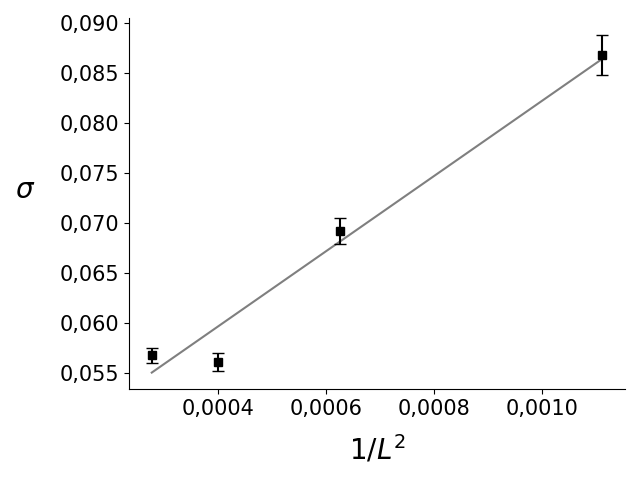}
\caption{Width $\sigma_{20}(L)$ of the variation $V_{CS}$ with error bars from Tables~\ref{table8} and \ref{table9} with the linear fit (solid line). Upper row represent data for the left peak in Figures~\ref{fig7} and bottom row represent data for the right peak in Figures~\ref{fig7}. Left row is the RD data set and right row is the MD data set. Data sets of PM-20 model, $q=20$. }
\label{fig9}
\end{figure}

First, $\sigma$ depends on the linear lattice size $L$ inversely proportional to $1/L^2$. This is illustrated in Figures~\ref{fig8} and \ref{fig9}, where $\sigma$  is plotted against $1/L^2$ and the solid lines correspond to the fits to data from the last columns in Tables~\ref{table6}-\ref{table9}. 

Second, we can calculate the ratio of the PM-10 peak width to the PM-20 peak width, namely $\sigma_{10}(L)$/$\sigma_{20}(L)$. To do this, we compute the ratio of the width $\sigma_{10}(L)$ for a particular lattice size $L$ from the last column of Table~\ref{table6} to the width $\sigma_{10}(L)$ for a particular lattice size $L$ from the last column of Table~\ref{table8} for the MD dataset. Similarly, we compute the ratio for  the RD dataset.  The resulting ratio is shown in the Table~\ref{table10}. The ratio of the specific thermal widths for PM-10 and PM-20 calculated by the~({\ref{eq:LK}) method is approximately 1.318. All ratios are of the order of one, from which we can conclude that the variation width of the $V_{CS}$ probability does depend on the number of spin components and the latent heat. What we have not been able to determine is the influence of other numbers characterizing a particular Potts model with a particular number of spin components. We mentioned in this subsection that a particular quantity can have different multipliers. There are many of them, as mentioned in~\cite{Lee-1991}. 

In any case, we verify that the probability width $V_{CS}$ does not contradict the general form~(\ref{eq:LK}).
}

\begin{table}
\centering
\begin{tabular}{|c|c||c|c|c|}
\hline
Dataset & \textbf{L} & $\mu$ & $e_o$ & $\sigma$ \\ \hline
\multirow{4}{*}{\textbf{MD}} & 30 & -1.66374(12) & \multirow{4}{*}{-1.664252} & 0.01308(14) \\
 & 40 & -1.66852(5) &  & 0.00920(6) \\
 & 50 & -1.66635(6) &  & 0.00727(8) \\
 & 60 & -1.66614(3) &  & 0.00606(4) \\ \hline

\multirow{4}{*}{\textbf{RD}} & 30 & -1.65281(17) & \multirow{4}{*}{-1.664252} &  0.02937(21) \\
 & 40 & -1.66555(10)&  & 0.01805(11)\\
 & 50 & -1.66668(6) &  & 0.01493(7) \\
 & 60 & -1.66233(4) &  & 0.01217(5) \\ \hline
\end{tabular}
\caption{Estimation of peak position $\mu$ and width $\sigma$ from the variation $V_{CS}$ near energy $e_o$, for PM-10 model.}
\label{table6}
\end{table}

\begin{table}
\centering
\begin{tabular}{|c|c||c|c|c|}
\hline
  Dataset & \textbf{L} & $\mu$ & $e_d$ & $\sigma$  \\
\hline
 \multirow{4}{*}{\textbf{MD}} & 30 & -0.99447(97) & \multirow{4}{*}{-0.968203} & 0.07163(138) \\
 & 40 & -0.99953(81) &  & 0.05723(87) \\
 & 50 & -0.97043(32) &  & 0.05268(60) \\
 & 60 & -0.97458(23) &  & 0.04555(37) \\ \hline
 \multirow{4}{*}{\textbf{RD}} & 30 & -0.95612(47) & \multirow{4}{*}{-0.968203} &  0.04773(67) \\
 & 40 & -0.97693(21) &  & 0.02918(27) \\
 & 50 & -0.95333(16) &  & 0.02437(21) \\
 & 60 & -0.97422(10) &  & 0.01893(12) \\ \hline
\end{tabular}
\caption{Estimation of peak position $\mu$ and width $\sigma$ from the variation $V_{CS}$ near energy $e_d$, for PM-10 model.}
\label{table7}
\end{table}

\begin{table}
\centering
\begin{tabular}{|c|c||c|c|c|}
\hline
     Dataset &  \textbf{L} &  $\mu$ & $e_o$ &  $\sigma$  \\
\hline
 \multirow{4}{*}{\textbf{MD}} & 30 & -1.81636(8) & \multirow{4}{*}{-1.820684} & 0.00926(9) \\
 & 40 & -1.82305(7) &  & 0.00636(9) \\
 & 50 & -1.82014(3) &  & 0.00474(4) \\
 & 60 & -1.81827(3) &  & 0.00393(4) \\ \hline

 \multirow{4}{*}{\textbf{RD}} & 30 & -1.81825(16) & \multirow{4}{*}{-1.820684} &  0.02153(20) \\
 & 40 & -1.81859(9) &  & 0.01606(11)\\
 & 50 & -1.81233(6) &  & 0.01290(7) \\
 & 60 & -1.82047(4) &  & 0.01003(5) \\ \hline
\end{tabular}
\caption{Estimation of peak position $\mu$ and width $\sigma$ from the variation $V_{CS}$ near energy $e_o$, for PM-20 model.}
\label{table8}
\end{table}

\begin{table}
\centering
\begin{tabular}{|c|c||c|c|c|}
\hline
     Dataset &  \textbf{L} &  $\mu$ & $e_d$ &  $\sigma$  \\
\hline
 \multirow{4}{*}{\textbf{MD}} & 30 & -0.66018(146) & \multirow{4}{*}{-0.626529} & 0.08682(202) \\
 & 40 & -0.65383(96) &  & 0.06924(130) \\
 & 50 & -0.66662(96) &  & 0.05614(90)  \\
 & 60 & -0.64900(58) &  & 0.05680(75)  \\ \hline

 \multirow{4}{*}{\textbf{RD}} & 30 & -0.61982(66) & \multirow{4}{*}{-0.626529} &  0.06752(118) \\
 & 40 & -0.65190(57) &  & 0.04730(75)  \\
 & 50 & -0.60936(24) &  & 0.03381(33)  \\
 & 60 & -0.63806(12) &  & 0.02692(16)  \\ \hline
\end{tabular}
\caption{ Estimation of peak position $\mu$ and width $\sigma$ from the variation $V_{CS}$ near energy $e_d$, for PM-20 model.}
\label{table9}
\end{table}

\begin{table}
\centering
\begin{tabular}{|c|c|c|c|}
\hline
 & \textbf{L} & left peak & right peak \\ \hline
 \multirow{4}{*}{\textbf{MD}} & 30 & 1.41 & 0.825 \\
 & 40 & 1.45 & 0.827 \\
 & 50 & 1.53 & 0.938 \\
 & 60 & 1.54 & 0.802 \\ \hline

 \multirow{4}{*}{\textbf{RD}} & 30 & 1.36 & 0.707 \\
 & 40 & 1.12 & 0.617 \\
 & 50 & 1.16 & 0.721 \\
 & 60 & 1.21 & 0.703 \\ \hline
\end{tabular}
\caption{Ratio of the peak width $\sigma_{10}(L)/\sigma_{20}(L)$ near the ordered energy $e_o$ and disordered energy $e_d$.}
\label{table10}
\end{table}

\subsection{Coexistence phase} 

 In contrast to binary classification in the case of a phase transition of the second order, our approach is based on ternary classification, using for training exactly known energies of the ordered phase $e_o$ and disordered phase $e_d$. To do this, we need to train the neural network on samples modeled at a certain value of energy $e$.  For this purpose, we use the microcanonical population annealing algorithm (MCPA)~\cite{Rose-2019,MS-2024,Fadeeva-2024}, which anneals a large population of the modeled system in energy space.  The trained network is used to classify all samples taken at a given energy $e$ into those that are more likely to be in ordered, disordered, or coexistence phase. Thus, we can estimate the probability that a given sample belongs to one of the three phases. This protocol allows us to restore critical energies and latent heat with reasonable accuracy.

\begin{figure}
\center
\includegraphics[width=1\linewidth]{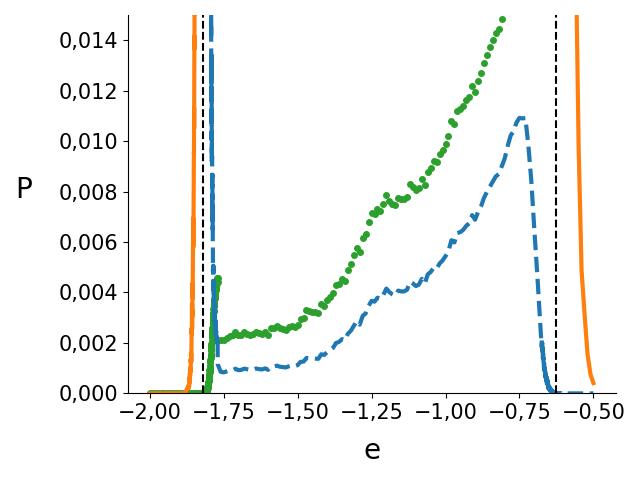}
\caption{Enlarged details of the left panel, last line of the Figure~\ref{fig5} showing small probability values with some features discussed in the Discussion section. Potts model with 20 components, PM-20 with grid size $L=60$, RD dataset.}
\label{fig10}
\end{figure} 

At the same time, we found that the estimated probabilities contain some information about the details of the coexistence phase. It is widely believed that there are four phase transitions in the coexistence phase, which is a random mixture of ordered and disordered phases~\cite{Neuhaus-2003,Rose-2019}. In the case of the Potts model, these are droplets that reflect the ordered and disordered phases~\cite{Zia-1990}. 

 In the Figure~\ref{fig10}, the solid orange line corresponds to the coexistence phase probability $P_{CS}$, the dotted green line corresponds to the disordered phase probability $P_{DS}$, and the dashed blue line corresponds to the ordered phase probability $P_{OS}$. The rightmost peak of $P_{OS}$ can be associated with the transition at energy $e_1(L)$ using the notations of Rose and Machta~\cite{Rose-2019}. It is associated with the fluctuating droplets of OS phase within the DS sea, which vanishes at $e_d$. It is the precursor of the phase transition from coexistence phase CS to disordered phase DS, as $e_1(L)\rightarrow e_d$ in the thermodynamic limit. By analogy, the leftmost peak of $P_{OD}$ can be associated with the transition at energy $e_4(L)$ and reflects DS droplets within the emerging sea of OS phase. Again, $e_4(L)\rightarrow e_o$ and  the DS droplets completely vanishes at $e_o$. So, the transitions at  $e_1(L)$ and $e_4(L)$ can be obtained only in the systems of finite size, and are not real phase transitions.
  
More interesting is the presence of small extrema on the DS and OS curves in the middle of the coexistence phase. They can be related to the wrapping clusters when DS or OS  droplets reach opposite boundaries of the system and due to the periodic boundaries will ``wrap'' around the torus. It is argued~\cite{Zia-1990,Neuhaus-2003,Rose-2019} that the  transitions at $e_2$ and $e_3<e_2$  associated with the wrapping droplet OS and wrapping droplet DS, respectively, exist in the thermodynamic limit. Note that the extrema shown in our Figure~\ref{fig10} are qualitatively similar to the extrema of a very different function, the integrated autocorrelation time shown in the Figure~5 of the article~\cite{Rose-2019}.   

It would be interesting to apply the phase probability estimation method proposed in the Letter to a regular analysis of the above pattern. With a possible demonstration of  regular limits of $e_1(L)$ and $e_4(L)$ with increasing $L$, and also a clear idea of droplet$\leftrightarrow$wrapping cluster transitions.

\section{Discussion}

The application of machine learning in statistical physics is developing intensively. The main objective is to investigate the details of applicability of this new research method. In other areas of physics, such as high-energy physics~\cite{HEP}, the use of a trained neural network model accelerates the pre-screening of uninteresting events in the analysis of particle scattering data. Training a neural network is a very labor-intensive process, but the trained network makes predictions at a much faster rate than other software models. 
Using this approach in statistical physics provides correct statistics  when generating statistically correct system configurations, for example, using diffusion networks~\cite{Marinari}.

In our proposed ternary sample classification method, we get the opportunity to investigate the probability of belonging to one of the three phases of the system state. We are not aware of any other approach that allows us to estimate such probabilities. As we noted above,our approach allows us to examine the details of the coexistence phase.

It would be interesting to apply the same  ternary classification approach to study phase transitions in non spin-system domain. For example, a fairly simple yet interesting problem with three phases is the behaviour of the hard disks in two-dimensions~\cite{Bernard-2011}. The solid, hexatic, and liquid phases are separated by a continuous transition and a first-order phase transition. It would be natural to use  the density values $\eta_{sh}=0.700$ and $\eta_{hl}=0.716$ from the paper~\cite{Bernard-2011}, which separates these phases, in the training process and use trained neural network model to predict the probabilities of all three phases.

\begin{acknowledgments}

The simulation was carried out using the high-performance computing resources of the National Research University Higher School of Economics.

The first short version of this article was supported by Russian Science Foundation grant No. 22-11-00259, and the final extended version was supported by grant No. 25-11-00158.

\end{acknowledgments}



\begin{thebibliography}{8}
\bibitem{footnote} Since the critical values of $e_o$ and $e_d$ are irrational and the energies have rational values on a finite size lattice, exact equalities are not achieved by numerical investigation.
\end{thebibliography}
\end{document}